\documentclass[journal]{IEEEtran}%
\usepackage{amsfonts}
\usepackage{amsmath}
\usepackage{amssymb}
\usepackage{graphicx}%
\providecommand{\U}[1]{\protect\rule{.1in}{.1in}}
\newtheorem{theorem}{Theorem}

\newtheorem{definition}[theorem]{Definition}

\newtheorem{lemma}[theorem]{Lemma}

\newtheorem{proposition}[theorem]{Proposition}

\let\originalleft\left
\let\originalright\right
\def\left#1{\mathopen{}\originalleft#1}
\def\right#1{\originalright#1\mathclose{}}

\begin{document}

\title{Polar codes for classical-quantum channels}
\author{Mark M. Wilde and Saikat Guha\thanks{Mark M. Wilde is with the School of
Computer Science, McGill University, Montreal, Quebec H3A 2A7, Canada. Saikat
Guha is with the Quantum Information Processing Group,
Raytheon BBN Technologies, Cambridge, Massachusetts, USA 02138. (E-mail: mark.wilde@mcgill.ca; sguha@bbn.com)}}
\maketitle

\begin{abstract}
Holevo, Schumacher, and Westmoreland's coding theorem guarantees the existence
of codes that are capacity-achieving for the task of sending classical data
over a channel with classical inputs and quantum outputs. Although they
demonstrated the existence of such codes, their proof does not provide an
explicit construction of codes for this task. The aim of the present paper is
to fill this gap by constructing near-explicit ``polar'' codes that are
capacity-achieving. The codes exploit the channel polarization phenomenon
observed by Arikan for the case of classical channels. Channel polarization is
an effect in which one can synthesize a set of channels, by ``channel
combining'' and ``channel splitting,'' in which a fraction of the synthesized
channels are perfect for data transmission while the other fraction are
completely useless for data transmission, with the good fraction equal to the
capacity of the channel. The channel polarization effect then leads to a
simple scheme for data transmission: send the information bits through the
perfect channels and ``frozen'' bits through the useless ones. The main
technical contributions of the present paper are threefold. First, we leverage several known
results from the quantum information literature to demonstrate that the
channel polarization effect occurs for channels with classical inputs and
quantum outputs. We then construct linear polar codes based on this effect,
and the encoding complexity is $O(N \log N)$, where $N$ is the blocklength of the code.
We also demonstrate that a quantum successive cancellation decoder works
well, in the sense that the word error rate decays exponentially with the blocklength of the code.
For this last result, we exploit Sen's recent ``non-commutative union bound'' that holds
for a sequence of projectors applied to a quantum state.
\end{abstract}

\section{Introduction}

Shannon's fundamental contribution was to establish the capacity of a noisy
channel as the highest rate at which a sender can reliably transmit data to a
receiver \cite{S48}. His method of proof exploited the probabilistic method
and was thus non-constructive. Ever since Shannon's contribution, researchers
have attempted to construct error-correcting codes that can reach the capacity
of a given channel. Some of the most successful schemes for error correction
are turbo codes and low-density parity-check codes~\cite{RU08}, with numerical
results demonstrating that these codes perform well for a variety of channels.
In spite of the success of these codes, there is no proof that they are
capacity achieving for channels other than the erasure channel \cite{K09}.

Recently, Arikan constructed polar codes and proved that they are capacity
achieving for a wide variety of channels \cite{A09}. Polar codes exploit the
phenomenon of \textit{channel polarization}, in which a simple, recursive
encoding synthesizes a set of channels that polarize, in the sense that a
fraction of them become perfect for transmission while the other fraction are
completely noisy and thus useless for transmission. The fraction of the
channels that become perfect for transmission is equal to the capacity of the
channel. In addition, the complexity of both the encoding and decoding scales
as $O\left(  N\log N\right)  $, where $N$ is the blocklength of the code.
Arikan developed polar codes after studying how the techniques of channel
combining and channel splitting affect the rate and reliability of a channel
\cite{A06}. Arikan and others have now extended the methods of polar coding to
many different settings, including arbitrary discrete memoryless channels
\cite{STA09}, source coding \cite{A10}, lossy source coding \cite{K09,KU10},
and the multiple access channel with two senders and one receiver \cite{STY10}.

All of the above results are important for determining both the limits on data transmission and methods for achieving these limits on classical channels. The description of a classical channel $p_{Y|X}$ arises from modeling the signaling alphabet, the physical transmission medium, and the receiver measurement. If we are interested in accurately evaluating and reaching the true data-transmission limits of the physical channels, with an unspecified receiver measurement, and whose information carriers require a quantum-mechanical description, then it becomes necessary to invoke the laws of quantum mechanics. Examples of such channels include deep-space optical channels and ultra-low-temperature quantum-noise-limited RF channels. Achieving the classical communication capacity for such (quantum) channels often requires making collective measurements at the receiver, an action for which no classical description or implementation exists.  The quantum-mechanical approach to information theory \cite{NC00,W11} is not merely a formality or technicality---encoding classical information with quantum states and decoding with collective measurements on the channel outputs \cite{Hol98,SW97} can dramatically improve data transmission rates, for example if the sender and receiver are operating in a low-power regime for a pure-loss optical channel (which is a practically relevant regime for long haul free-space terrestrial and deep-space optical communication) \cite{GGLMSY04,G11}. Also, encoding with entangled inputs to the channels can increase capacity for certain channels \cite{H09}, a superadditive effect which simply does not occur for classical channels.

The proof of one of the most important theorems of quantum information theory
is due to Holevo~\cite{Hol98}, Schumacher, and Westmoreland~\cite{SW97} (HSW).
They showed that the Holevo information of a quantum channel is an achievable
rate for classical communication over it. Their proof of the HSW\ theorem
bears some similarities with Shannon's technique (including the use of random
coding), but their main contribution was the construction of a quantum
measurement at the receiving end that allows for reliable decoding at the
Holevo information rate. Since the proof of the HSW\ theorem, several
researchers have improved the proof's error analysis \cite{HN03}, and others
have demonstrated different techniques for achieving the Holevo information
\cite{Hol98a,W99,DD06,ON07,PhysRevLett.108.200501}. Very recently, Giovannetti \textit{et al}.~proved
that a sequential decoding approach can achieve the Holevo information
\cite{GLM10}.\ The sequential decoding approach has the receiver ask, through
a series of dichotomic quantum measurements, whether the output of the channel
was the first codeword, the second codeword, etc.~(this approach is similar in
spirit to a classical \textquotedblleft jointly-typical\textquotedblright%
\ decoder \cite{book1991cover}). As long as the rate of the code is less than
the Holevo information, then this sequential decoder will correctly identify
the transmitted codeword with asymptotically negligible error probability. Sen
recently simplified the error analysis of this sequential decoding approach
(rather significantly) by introducing a \textquotedblleft non-commutative
union bound\textquotedblright\ in order to bound the error probability of
quantum sequential decoding \cite{S11}.

In spite of the large amount of effort placed on proving that the Holevo
information is achievable, there has been relatively little work on devising
explicit codes that approach the Holevo information rate.\footnote{This is
likely due to the large amount of effort that the quantum information
community has put towards \textit{quantum} error correction~\cite{DNM09},
which is important for the task of transmitting quantum bits over a noisy
quantum channel or for building a fault-tolerant quantum computer.
Also, there might be a general belief that classical coding strategies would
extend easily for sending classical information over quantum channels, but
this is not the case given that collective measurements on channel outputs are
required to achieve the Holevo information rate and the classical strategies
do not incorporate these collective measurements.} The aim of the present
paper is to fill this gap by generalizing the polar coding approach to quantum
channels. In doing so, we construct the first explicit class of linear codes
that approach the Holevo information rate with asymptotically small error probability.

The main technical contributions of the present paper are as follows:

\begin{enumerate}
\item We characterize \textit{rate} with the symmetric Holevo information
\cite{Holevo73,Hol98,NC00,W11} and \textit{reliability} with the fidelity
\cite{U73,J94,NC00,W11} between channel outputs corresponding to different
classical inputs. These parameters generalize the symmetric Shannon capacity and the
Bhattacharya parameter~\cite{A09}, respectively, to the quantum case. We
demonstrate that the symmetric Holevo information and the fidelity polarize
under a recursive channel transformation similar to Arikan's \cite{A09}, by
exploiting Arikan's proof ideas \cite{A09} and several tools from the quantum
information literature \cite{LR73,FvG99,H00,NC00,RFZ10,W11}.

\item The second contribution of ours is the generalization of
Arikan's successive cancellation decoder \cite{A09} to the quantum case. We
exploit ideas from quantum hypothesis testing \cite{H69,Hol72,Hel76,H06,NH07}
in order to construct the quantum successive cancellation decoder, and we use
Sen's recent \textquotedblleft non-commutative union bound\textquotedblright%
\ \cite{S11}\ in order to demonstrate that the decoder performs reliably in
the limit of many channel uses, while achieving the symmetric Holevo
information rate.
\end{enumerate}

The complexity of the encoding part of our polar coding scheme is $O\left(
N\log N\right)  $ where $N$ is the blocklength of the code (the argument for
this follows directly from Arikan's \cite{A09}). However, we have not yet been
able to show that the complexity of the decoding part is $O\left(  N\log N\right)
$ (as is the case with Arikan's decoder \cite{A09}). Determining how to
simplify the complexity of the decoding part is the subject of ongoing
research. For now, we should regard our contribution in this paper as a more
explicit method for achieving the Holevo information rate (as compared to
those from prior work~\cite{Hol98,SW97,Hol98a,W99,DD06,ON07}).

One might naively think from a casual glance at our paper that Arikan's
results \cite{A09} directly apply to our quantum scenario here, but this is
not the case. If one were to impose single-symbol detection on the outputs of
the quantum channels,\footnote{For instance, all known conventional optical
receivers are single-symbol detectors. They detect each modulated pulse
individually, followed by classical postprocessing.} such a procedure
would induce a classical channel from
input to output. In this
case, Arikan's results do apply in that they can attain the Shannon capacity
of this induced classical channel.

However, the Shannon capacity of the best
single-symbol detection strategy may be far below the Holevo limit \cite{GGLMSY04,G11}.
Attaining the Holevo information rate generally requires the receiver to perform
collective measurements (physical detection of the quantum state of the entire
codeword that may not be realizable by detecting single symbols one at a time). 
We should stress that what we are doing in this paper is different from a naive application
of Arikan's results. First, our polar coding rule
depends on a quantum parameter, the fidelity, rather than the Bhattacharya
distance (a classical parameter). The polar coding rule is then different
from Arikan's, and we would thus expect a larger fraction of the channels to
be ``good'' channels than if one were to impose a single-symbol measurement
and exploit Arikan's polar coding rule with the Bhattacharya distance. Second,
the quantum measurements in our quantum successive cancellation decoder are
collective measurements performed on all of the channel outputs. Were it not
so, then our polar coding scheme would not achieve the Holevo information rate in general.

We organize the rest of the paper as follows. The next section provides an overview of polar coding for
classical-quantum channels (channels with classical inputs and quantum
outputs). This overview states the main concepts and the important theorems,
while saving their proofs for later in the paper. The main concepts include
channel combining, channel splitting, channel polarization, rate of
polarization, quantum successive cancellation decoding, and polar code
performance. Section~\ref{sec:rec-chan-trans} gives more detail on how
recursive channel combining and splitting lead to transformation of rate and
reliability in the direction of polarization.
Section~\ref{sec:channel-polarization}\ proves that channel polarization
occurs under the transformations given in Section~\ref{sec:rec-chan-trans}
(the proofs in Section~\ref{sec:channel-polarization} are identical to
Arikan's \cite{A09} because they merely exploit his martingale approach). We
prove in Section~\ref{sec:succ-canc-analysis} that the performance of the
polar coding scheme is good, by analyzing the error probability under quantum
successive cancellation decoding. We finally conclude in
Section~\ref{sec:concl}\ with a summary and some open questions.

\section{Overview of Results}

Our setting involves a classical-quantum channel $W$ with a classical input
$x$ and a quantum output $\rho_{x}$:%
\[
W:x\rightarrow\rho_{x},
\]
where $x\in\left\{  0,1\right\}  $ and $\rho_{x}$ is a unit trace, positive
operator called a \textit{density operator}. We can associate a probability
distribution and a classical label with the states $\rho_{0}$ and $\rho_{1}$
by writing the following \textit{classical-quantum state} \cite{W11}:%
\[
\rho^{XB}\equiv\frac{1}{2}\left\vert 0\right\rangle \left\langle 0\right\vert
^{X}\otimes\rho_{0}^{B}+\frac{1}{2}\left\vert 1\right\rangle \left\langle
1\right\vert ^{X}\otimes\rho_{1}^{B}.
\]

Two important parameters for characterizing any classical-quantum channel are
its rate and reliability.\footnote{We are using the same terminology as Arikan
\cite{A09}.} We define the rate in terms of the channel's symmetric Holevo
information $I\left(  W\right)  $ where%
\[
I\left(  W\right)  \equiv I\left(  X;B\right)  _{\rho}.
\]
$I\left(  X;B\right)  _{\rho}$ is the quantum mutual information of the state
$\rho^{XB}$, defined as%
\[
I\left(  X;B\right)  _{\rho}\equiv H\left(  X\right)  _{\rho}+H\left(
B\right)  _{\rho}-H\left(  XB\right)  _{\rho},
\]
and the von Neumann entropy $H\left(  \sigma\right)  $ of any density operator
$\sigma$ is defined as%
\[
H\left(  \sigma\right)  \equiv-\text{Tr}\left\{  \sigma\log_{2}\sigma\right\}
.
\]
(Observe that the von Neumann entropy of $\sigma$ is equal to the Shannon entropy of its eigenvalues.)
It is also straightforward to verify that
\[
I\left(  W\right)  = H((\rho_{0}^{B} + \rho_{1}^{B})/2) - %
H(\rho_{0}^{B})/2 - H(\rho_{1}^{B})/2 .
\]
The symmetric Holevo information is non-negative by concavity of von Neumann
entropy, and it can never exceed one if the system $X$ is a classical binary
system (as is the case for the classical-quantum state $\rho^{XB}$).
Additionally, the symmetric Holevo information is equal to zero if there is no
correlation between $X$ and $B$. It is equal to the capacity of the channel
$W$ for transmitting classical bits over it if the input prior distribution is
restricted to be uniform \cite{Hol98,SW97}. It also generalizes the symmetric
capacity \cite{A09} to the quantum setting given above.

We define the reliability of the channel $W$ as the fidelity between the
states $\rho_{0}$ and $\rho_{1}$ \cite{U73,J94,NC00,W11}:%
\[
F\left(  \rho_{0},\rho_{1}\right)  \equiv\left\Vert \sqrt{\rho_{0}}\sqrt
{\rho_{1}}\right\Vert _{1}^{2},
\]
where $\left\Vert A\right\Vert _{1}$ is the nuclear norm of the operator $A$:%
\[
\left\Vert A\right\Vert _{1}=\text{Tr}\left\{  \sqrt{A^{\dag}A}\right\}  .
\]
Let $F\left(  W\right)  $ denote the reliability of the channel $W$:%
\[
F\left(  W\right)  \equiv F\left(  \rho_{0},\rho_{1}\right)  .
\]
The fidelity is equal to a number between zero and one, and it characterizes
how \textquotedblleft close\textquotedblright\ two quantum states are to one
another. It is equal to zero if and only if there exists a quantum measurement
that can perfectly distinguish the states, and it is equal to one if the
states are indistinguishable by any measurement~\cite{NC00,W11}. The fidelity
generalizes the Bhattacharya parameter used in the classical
setting~\cite{A09}. Naturally, we would expect the channel $W$ to be perfectly
reliable if $F\left(  W\right)  =0$ and completely unreliable if $F\left(
W\right)  =1$. The fidelity also serves as a coarse bound on the probability
of error in discriminating the states $\rho_{0}$ and $\rho_{1}$
\cite{H06,CMMAB08}.

We would expect the symmetric Holevo information $I\left(  W\right)  \approx1$
if and only if the channel's fidelity $F\left(  W\right)  \approx0$ and vice
versa: $I\left(  W\right)  \approx0\Leftrightarrow F\left(  W\right)
\approx1$. The following proposition makes this intuition rigorous, and it
serves as a generalization of Arikan's first proposition regarding the
relationship between rate and reliability. We provide its proof in the appendix.

\begin{proposition}
\label{prop:IvsF}For any binary input classical-quantum channel of the above
form, the following bounds hold%
\begin{align}
I\left(  W\right)   &  \geq\log_{2}\left(  \frac{2}{1+\sqrt{F\left(  W\right)
}}\right)  ,\label{eq:rate-rel-1st-bound}\\
I\left(  W\right)   &  \leq\sqrt{1-F\left(  W\right)  }.
\label{eq:rate-rel-2nd-bound}%
\end{align}

\end{proposition}

\subsection{Channel Polarization}

The channel polarization phenomenon occurs after synthesizing a set of $N$
classical-quantum channels $\{W_{N}^{\left(  i\right)  }:1\leq i\leq N\}$ from
$N$ independent copies of the classical-quantum channel $W$. The effect is
known as \textquotedblleft polarization\textquotedblright\ because a fraction
of the channels $W_{N}^{\left(  i\right)  }$ become perfect for data
transmission,\footnote{One cannot expect to transmit more than one classical
bit over a perfect qubit channel due to Holevo's bound \cite{Holevo73}.} in
the sense that $I(W_{N}^{\left(  i\right)  })\approx1$ for the channels in
this fraction, while the channels in the complementary fraction become
completely useless in the sense that $I(W_{N}^{\left(  i\right)  })\approx0$
in the limit as $N$ becomes large. Also, the fraction of channels that do not
exhibit polarization vanishes as $N$ becomes large. One can induce the
polarization effect by means of channel combining and channel splitting.

\subsubsection{Channel Combining}

\begin{figure}[ptb]
\begin{center}
\includegraphics[
width=1.8343in
]{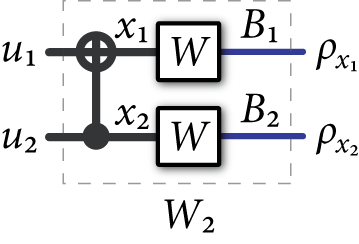}
\end{center}
\caption{The channel $W_{2}$ synthesized from the first level of recursion.
Thick lines denote classical systems while thin lines denote quantum systems
(this is our convention for the other figures as well).
The depicted gate acting on the channel input is a classical controlled-NOT (CNOT) gate, where the
filled-in circle acts on the source bit and the other circle acts on the
target bit. Its truth table is $\left(  u_{1},u_{2}\right)  \rightarrow\left(
u_{1}\oplus u_{2},u_{2}\right)  $.}%
\label{fig:first-recursion}%
\end{figure}The channel combining phase takes copies of a classical-quantum
channel $W$ and builds from them an $N$-fold classical-quantum channel $W_{N}$
in a recursive way, where $N$ is any power of two:$\ N=2^{n}$, $n\geq0$. The
zeroth level of recursion merely sets $W_{1}\equiv W$. The first level of
recursion combines two copies of $W_{1}$ and produces the channel $W_{2}$,
defined as%
\begin{equation}
W_{2}:u_{1}u_{2}\rightarrow W_{2}^{B_{1}B_{2}}\left(  u_{1},u_{2}\right)  ,
\label{eq:channel-combine-1}%
\end{equation}
where%
\[
W_{2}^{B_{1}B_{2}}\left(  u_{1},u_{2}\right)  \equiv\rho_{u_{1}\oplus u_{2}%
}^{B_{1}}\otimes\rho_{u_{2}}^{B_{2}}.
\]
Figure~\ref{fig:first-recursion} depicts this first level of recursion.

\begin{figure}[ptb]
\begin{center}
\includegraphics[
width=2.437in
]{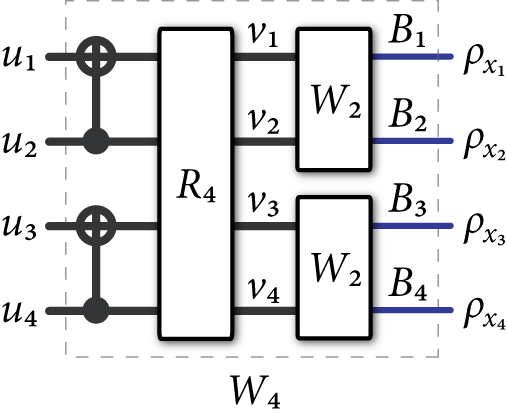}
\end{center}
\caption{The second level of recursion in the channel combining phase.}%
\label{fig:second-recursion}%
\end{figure}

The second level of recursion takes two copies of $W_{2}$ and produces the
channel $W_{4}$:%
\begin{equation}
W_{4}:u_{1}u_{2}u_{3}u_{4}\rightarrow W_{4}^{B_{1}B_{2}B_{3}B_{4}}\left(
u_{1},u_{2},u_{3},u_{4}\right)  ,\label{eq:channel-combine-2}%
\end{equation}
where%
\begin{multline*}
W_{4}^{B_{1}B_{2}B_{3}B_{4}}\left(  u_{1},u_{2},u_{3},u_{4}\right)  \\
\equiv W_{2}^{B_{1}B_{2}}\left(  u_{1}\oplus u_{2},u_{3}\oplus u_{4}\right)
\otimes W_{2}^{B_{3}B_{4}}\left(  u_{2},u_{4}\right)  ,
\end{multline*}
so that%
\begin{multline*}
W_{4}^{B_{1}B_{2}B_{3}B_{4}}\left(  u_{1},u_{2},u_{3},u_{4}\right)  \\
=\rho_{u_{1}\oplus u_{2}\oplus u_{3}\oplus u_{4}}^{B_{1}}\otimes\rho
_{u_{3}\oplus u_{4}}^{B_{2}}\otimes\rho_{u_{2}\oplus u_{4}}^{B_{3}}\otimes
\rho_{u_{4}}^{B_{4}}.
\end{multline*}
Figure~\ref{fig:second-recursion}\ depicts the second level of recursion.

The operation $R_{4}$ in Figure~\ref{fig:second-recursion}\ is a permutation
that takes $\left(  u_{1},u_{2},u_{3},u_{4}\right)  \rightarrow\left(
u_{1},u_{3},u_{2},u_{4}\right)  $. One can then readily check that the mapping
from the row vector $u_{1}^{4}$ to the channel inputs $x_{1}^{4}$ is a linear
map given by $x_{1}^{4}=u_{1}^{4}G_{4}$ with%
\[
G_{4}\equiv%
\begin{bmatrix}
1 & 0 & 0 & 0\\
1 & 0 & 1 & 0\\
1 & 1 & 0 & 0\\
1 & 1 & 1 & 1
\end{bmatrix}
.
\]

The general recursion at the $n^{\text{th}}$ level is to take two copies of
$W_{N/2}$ and synthesize a channel $W_{N}$ from them. The first part is to
transform the input sequence $u^{N}$ according to the following rule for all
$i\in\left\{  1,\ldots,N/2\right\}  $:%
\begin{align*}
s_{2i-1} &  =u_{2i-1}\oplus u_{2i},\\
s_{2i} &  =u_{2i}.
\end{align*}
The next part of the transformation is a \textquotedblleft reverse
shuffle\textquotedblright\ $R_{N}$ that performs the transformation:%
\begin{multline*}
\left(  s_{1},s_{2},s_{3},s_{4},\ldots,s_{N-1},s_{N}\right)  \\
\rightarrow\left(  s_{1},s_{3},\ldots,s_{N-1},s_{2},s_{4},\ldots,s_{N}\right)
.
\end{multline*}
The resulting bit sequence is the input to the two copies of $W_{N/2}$.

The overall transformation on the input sequence $u^{N}$ is a linear
transformation given by $x^{N}=u^{N}G_{N}$ where%
\begin{equation}
G_{N}=B_{N}F^{\otimes n}, \label{eq:GN-matrix}%
\end{equation}
where%
\[
F\equiv%
\begin{bmatrix}
1 & 0\\
1 & 1
\end{bmatrix}
,
\]
and $B_{N}$ is a permutation matrix known as a ``bit-reversal'' operation \cite{A09}.

\subsubsection{Channel Splitting}

The channel splitting phase consists of taking the channels $W_{N}$ induced by
the transformation $G_{N}$ and defining new channels $W_{N}^{\left(  i\right)
}$ from them. Let $\rho_{u^{N}}$ denote the output of the channel $W_{N}$ when
inputting the bit sequence $u^{N}$. We define the $i^{\text{th}}$ split
channel $W_{N}^{\left(  i\right)  }$ as follows:%
\begin{equation}
W_{N}^{\left(  i\right)  }:u_{i}\rightarrow\rho_{\left(  i\right)  ,u_{i}%
}^{U_{1}^{i-1}B^{N}}, \label{eq:split-channels}%
\end{equation}
where%
\begin{align}
\rho_{\left(  i\right)  ,u_{i}}^{U_{1}^{i-1}B^{N}}  &  \equiv\sum_{u_{1}%
^{i-1}}\frac{1}{2^{i-1}}\left\vert u_{1}^{i-1}\right\rangle \left\langle
u_{1}^{i-1}\right\vert ^{U_{1}^{i-1}}\otimes\overline{\rho}_{u_{1}^{i}}%
^{B^{N}},\\
\overline{\rho}_{u_{1}^{i}}^{B^{N}}  &  \equiv\sum_{u_{i+1}^{N}}\frac
{1}{2^{N-i}}\rho_{u^{N}}^{B^{N}}. \label{eq:averaged-cond-states}%
\end{align}
We can also write as an alternate notation
\[
W_{N}^{\left(  i\right)  }(u_{i}) = \rho_{\left(  i\right)  ,u_{i}}%
^{U_{1}^{i-1}B^{N}}.
\]
These channels have the same interpretation as Arikan's split channels
\cite{A09}---they are the channels induced by a \textquotedblleft
genie-aided\textquotedblright\ quantum successive cancellation decoder, in
which the $i^{\text{th}}$ decision measurement estimates $u_{i}$ given that
the channel output $\rho_{u^{N}}^{B^{N}}$ is available, after observing the
previous bits $u_{1}^{i-1}$ correctly, and if the distribution over $u_{i+1}^{N}$ is
uniform. These split channels arise in our analysis of the error probability
for quantum successive cancellation decoding.

\subsubsection{Channel Polarization}

Our channel polarization theorem below is similar to Arikan's Theorem~1
\cite{A09}, though ours applies for classical-quantum channels with
binary inputs and quantum outputs:

\begin{theorem}
[Channel Polarization]\label{thm:channel-polarization}The classical-quantum
channels $W_{N}^{\left(  i\right)  }$ synthesized from the channel $W^{\otimes
N}$ polarize, in the sense that the fraction of indices $i\in\left\{
1,\ldots,N\right\}  $ for which $I(W_{N}^{\left(  i\right)  })\in(1-\delta,1]$
goes to the symmetric Holevo information $I\left(  W\right)  $ and the
fraction for which $I(W_{N}^{\left(  i\right)  })\in\lbrack0,\delta)$ goes to
$1-I\left(  W\right)  $ for any $\delta\in\left(  0,1\right)  $ as $N$ goes to
infinity through powers of two.
\end{theorem}

The proof of the above theorem is identical to Arikan's proof with a
martingale approach \cite{A09}. For completeness, we provide a brief proof in
Section~\ref{sec:channel-polarization}.

\subsubsection{Rate of Polarization}

It is important to characterize the speed with which the polarization
phenomenon comes into play for the purpose of proving this paper's polar
coding theorem. We exploit the fidelity $F(W^{\left(  i\right)  })$ of the
split channels in order to characterize the rate of polarization:%
\begin{equation}
F(W_{N}^{\left(  i\right)  })\equiv F(\rho_{\left(  i\right)  ,0}^{U_{1}%
^{i-1}B^{N}},\rho_{\left(  i\right)  ,1}^{U_{1}^{i-1}B^{N}}).
\label{eq:channel-fidelity}%
\end{equation}
The theorem below exploits the exponential convergence results of Arikan and
Telatar \cite{AT09}, which improved upon Arikan's original convergence results
\cite{A09} (note that we could also use the more general results in
Ref.~\cite{KSU10}):

\begin{theorem}
[Rate of Polarization]\label{thm:polar-rate}Given any classical-quantum
channel $W$ with $I\left(  W\right)  >0$, any $R<I\left(  W\right)  $, and any
constant $\beta<1/2$, there exists a sequence of sets $\mathcal{A}_{N}%
\subset\left\{  1,\ldots,N\right\}  $ with $\left\vert \mathcal{A}%
_{N}\right\vert \geq NR$ such that%
\[
\sum_{i\in\mathcal{A}_{N}}\sqrt{F(W_{N}^{\left(  i\right)  })}=o(2^{-N^{\beta
}}).
\]
Conversely, suppose that $R>0$ and $\beta>1/2$. Then for any sequence of sets
$\mathcal{A}_{N}\subset\left\{  1,\ldots,N\right\}  $ with $\left\vert
\mathcal{A}_{N}\right\vert \geq NR$, the following result holds%
\[
\max\left\{  \sqrt{F(W_{N}^{\left(  i\right)  })}:i\in\mathcal{A}_{N}\right\}
=\omega(2^{-N^{\beta}}).
\]

\end{theorem}

The proof of this theorem exploits our results in
Section~\ref{sec:rec-chan-trans}\ and Theorem~1 of Ref.~\cite{AT09}.

\subsection{Polar Coding}

The idea behind polar coding is to exploit the polarization effect for the
construction of a capacity-achieving code. The sender should transmit the information bits only
through the split channels $W_{N}^{\left(  i\right)  }$ for which the
reliability parameter $F(W_{N}^{\left(  i\right)  })$ is close to zero. In
doing so, the sender and receiver can achieve the symmetric Holevo information
$I\left(  W\right)  $ of the channel $W$.

\subsubsection{Coset Codes}

Polar codes arise from a special class of codes that Arikan calls
\textquotedblleft$G_{N}$-coset codes\textquotedblright\ \cite{A09}. These
$G_{N}$-coset codes are given by the following mapping from the input sequence
$u^{N}$ to the channel input sequence $x^{N}$:%
\[
x^{N}=u^{N}G_{N},
\]
where $G_{N}$ is the encoding matrix defined in (\ref{eq:GN-matrix}). Suppose
that $\mathcal{A}$ is some subset of $\left\{  1,\ldots,N\right\}  $. Then we
can write the above transformation as follows:%
\begin{equation}
x^{N}=u_{\mathcal{A}}G_{N}\left(  \mathcal{A}\right)  \oplus u_{\mathcal{A}%
^{c}}G_{N}\left(  \mathcal{A}^{c}\right)  , \label{eq:coset-code-decomp}%
\end{equation}
where $G_{N}\left(  \mathcal{A}\right)  $ denotes the submatrix of $G_{N}$
constructed from the rows of $G_{N}$ with indices in $\mathcal{A}$ and
$\oplus$ denotes vector binary addition.

Suppose that we fix the set $\mathcal{A}$ and the bit sequence $u_{\mathcal{A}%
^{c}}$. The mapping in (\ref{eq:coset-code-decomp}) then specifies a
transformation from the bit sequence $u_{\mathcal{A}}$ to the channel input
sequence $x^{N}$. This mapping is equivalent to a linear encoding for a code
that Arikan calls a $G_{N}$-coset code where the sequence $u_{\mathcal{A}^{c}%
}G_{N}\left(  \mathcal{A}^{c}\right)  $ identifies the coset. We can fully
specify a coset code by the parameter vector $\left(  N,K,\mathcal{A}%
,u_{\mathcal{A}^{c}}\right)  $ where $N$ is the length of the code,
$K=\left\vert \mathcal{A}\right\vert $ is the number of information bits,
$\mathcal{A}$ is a set that identifies the indices for the information bits,
and $u_{\mathcal{A}^{c}}$ is the vector of frozen bits. The polar coding rule
specifies a way to choose the indices for the information bits based on the
channel over which the sender is transmitting data.

\subsubsection{A Quantum Successive Cancellation Decoder}

The specification of the quantum successive cancellation decoder is what
mainly distinguishes Arikan's polar codes for classical channels from ours
developed here for classical-quantum channels. Let us begin with a $G_{N}%
$-coset code with parameter vector $\left(  N,K,\mathcal{A},u_{\mathcal{A}%
^{c}}\right)  $. The sender encodes the information bit vector $u_{\mathcal{A}%
}$ along with the frozen vector $u_{\mathcal{A}^{c}}$ according to the
transformation in (\ref{eq:coset-code-decomp}). The sender then transmits the
encoded sequence $x^{N}$ through the classical-quantum channel, leading to a
state $\rho_{x_{1}}\otimes\cdots\otimes\rho_{x_{N}}$, which is equivalent to a
state $\rho_{u^{N}}$ up to the transformation $G_{N}$. It is then the goal of
the receiver to perform a sequence of quantum measurements on the state
$\rho_{u^{N}}$ in order to determine the bit sequence $u^{N}$. We are assuming
that the receiver has full knowledge of the frozen vector $u_{\mathcal{A}^{c}%
}$ so that he does not make mistakes when decoding these bits.

Corresponding to the split channels $W_{N}^{\left(  i\right)  }$ in
(\ref{eq:split-channels}) are the following projectors that can attempt to
decide whether the input of the $i^{\text{th}}$ split channel is zero or one:%
\begin{align*}
\Pi_{\left(  i\right)  ,0}^{U_{1}^{i-1}B^{N}} &  \equiv\left\{  \sqrt
{\rho_{\left(  i\right)  ,0}^{U_{1}^{i-1}B^{N}}}-\sqrt{\rho_{\left(  i\right)
,1}^{U_{1}^{i-1}B^{N}}}\geq0\right\}  ,\\
\Pi_{\left(  i\right)  ,1}^{U_{1}^{i-1}B^{N}} &  \equiv I-\Pi_{\left(
i\right)  ,0}^{U_{1}^{i-1}B^{N}}\\
&  =\left\{  \sqrt{\rho_{\left(  i\right)  ,0}^{U_{1}^{i-1}B^{N}}}-\sqrt
{\rho_{\left(  i\right)  ,1}^{U_{1}^{i-1}B^{N}}}<0\right\}  ,
\end{align*}
where $\sqrt{A}$ denotes the square root of a positive operator $A$, $\left\{
B\geq0\right\}  $ denotes the projector onto the positive eigenspace of a
Hermitian operator~$B$, and $\left\{  B<0\right\}  $ denotes the projection
onto the negative eigenspace of~$B$. After some calculations, we can readily
see that%
\begin{align}
\Pi_{\left(  i\right)  ,0}^{U_{1}^{i-1}B^{N}} &  =\sum_{u_{1}^{i-1}}\left\vert
u_{1}^{i-1}\right\rangle \left\langle u_{1}^{i-1}\right\vert ^{U_{1}^{i-1}%
}\otimes\Pi_{\left(  i\right)  ,u_{1}^{i-1}0}^{B^{N}}%
,\label{eq:projectors-expanded-1}\\
\Pi_{\left(  i\right)  ,1}^{U_{1}^{i-1}B^{N}} &  =\sum_{u_{1}^{i-1}}\left\vert
u_{1}^{i-1}\right\rangle \left\langle u_{1}^{i-1}\right\vert ^{U_{1}^{i-1}%
}\otimes\Pi_{\left(  i\right)  ,u_{1}^{i-1}1}^{B^{N}}%
,\label{eq:projectors-expanded-2}%
\end{align}
where%
\begin{align*}
\Pi_{\left(  i\right)  ,u_{1}^{i-1}0}^{B^{N}} &  \equiv\left\{  \sqrt
{\overline{\rho}_{u_{1}^{i-1}0}^{B^{N}}}-\sqrt{\overline{\rho}_{u_{1}^{i-1}%
1}^{B^{N}}}\geq0\right\}  ,\\
\Pi_{\left(  i\right)  ,u_{1}^{i-1}1}^{B^{N}} &  \equiv\left\{  \sqrt
{\overline{\rho}_{u_{1}^{i-1}0}^{B^{N}}}-\sqrt{\overline{\rho}_{u_{1}^{i-1}%
1}^{B^{N}}}<0\right\}  .
\end{align*}

The above observations lead to a method for a successive cancellation decoder
similar to Arikan's~\cite{A09}, with the following decoding rule:%
\[
\hat{u}_{i}=\left\{
\begin{array}
[c]{cc}%
u_{i} & \text{if }i\in\mathcal{A}^{c}\\
h\left(  \hat{u}_{1}^{i-1}\right)   & \text{if }i\in\mathcal{A}%
\end{array}
\right.  ,
\]
where $h\left(  \hat{u}_{1}^{i-1}\right)  $ is the outcome of the following
$i^{\text{th}}$ measurement on the output of the channel (after $i-1$
measurements have already been performed):%
\[
\left\{  \Pi_{\left(  i\right)  ,\hat{u}_{1}^{i-1}0}^{B^{N}},\Pi_{\left(
i\right)  ,\hat{u}_{1}^{i-1}1}^{B^{N}}\right\}  .
\]
We are assuming that the measurement device outputs \textquotedblleft%
0\textquotedblright\ if the outcome $\Pi_{\left(  i\right)  ,\hat{u}_{1}%
^{i-1}0}^{B^{N}}$ occurs and it outputs \textquotedblleft1\textquotedblright%
\ otherwise. (Note that we can set $\Pi_{\left(  i\right)  ,\hat{u}_{1}%
^{i-1}u_{i}}^{B^{N}}=I$ if the bit $u_{i}$ is a frozen bit.) The above
sequence of measurements for the whole bit stream $u^{N}$ corresponds to a
positive operator-valued measure (POVM)~$\left\{  \Lambda_{u^{N}}\right\}  $
where%
\begin{multline*}
\Lambda_{u^{N}}\equiv\Pi_{\left(  1\right)  ,u_{1}}^{B^{N}}\cdots\Pi_{\left(
i\right)  ,u_{1}^{i-1}u_{i}}^{B^{N}}\cdots\\
\cdots\Pi_{\left(  N\right)  ,u_{1}^{N-1}u_{N}}^{B^{N}}\cdots\Pi_{\left(
i\right)  ,u_{1}^{i-1}u_{i}}^{B^{N}}\cdots\Pi_{\left(  1\right)  ,u_{1}%
}^{B^{N}},
\end{multline*}%
\[
\sum_{u_{\mathcal{A}}}\Lambda_{u^{N}}=I^{B^{N}}.
\]

The above decoding strategy is suboptimal in two regards. First, the decoder
assumes that the future bits are unknown (and random) even if the receiver has
full knowledge of the future frozen bits (this suboptimality is similar to the
suboptimality of Arikan's decoder~\cite{A09}). Second, the measurement
operators for making a decision are suboptimal as well because we choose them
to be projectors onto the positive eigenspace of the difference of the
\textit{square roots} of two density operators. The optimal bitwise decision
rule is to choose these operators to be the Helstrom-Holevo projector onto the
positive eigenspace of the difference of two density operators
\cite{H69,Hol72}. Having our quantum successive cancellation
decoder operate in these two different suboptimal ways allows for
us to analyze its performance easily (though, note that we could just as well have used
Helstrom-Holevo measurements to obtain bounds on the error probability). This suboptimality is
asymptotically negligible because the symmetric Holevo information is still an
achievable rate for data transmission even for the above choice of measurement operators.

\subsubsection{Polar Code Performance}

The probability of error $P_{e}\left(  N,K,\mathcal{A},u_{\mathcal{A}^{c}%
}\right)  $\ for code length $N$, number $K$ of information bits, set
$\mathcal{A}$ of information bits, and choice $u_{\mathcal{A}^{c}}$ for the
frozen bits is as follows:%
\begin{align*}
&  P_{e}\left(  N,K,\mathcal{A},u_{\mathcal{A}^{c}}\right)  \\
&  =\frac{1}{2^{K}}\sum_{u_{\mathcal{A}}}\text{Tr}\left\{  \left(
I-\Lambda_{u^{N}}\right)  \rho_{u^{N}}\right\}  \\
&  =1-\frac{1}{2^{K}}\sum_{u_{\mathcal{A}}}\text{Tr}\left\{  \Lambda_{u^{N}%
}\rho_{u^{N}}\right\}  \\
&  =1-\frac{1}{2^{K}}\sum_{u_{\mathcal{A}}}\text{Tr}\bigg\{\Pi_{\left(  N\right)
,u_{1}^{N-1}u_{N}}^{B^{N}}\cdots\Pi_{\left(  i\right)  ,u_{1}^{i-1}u_{i}%
}^{B^{N}}\cdots\\
&  \ \ \ \cdots\Pi_{\left(  1\right)  ,u_{1}}^{B^{N}}\ \rho_{u^{N}}%
\ \Pi_{\left(  1\right)  ,u_{1}}^{B^{N}}\cdots\Pi_{\left(  i\right)
,u_{1}^{i-1}u_{i}}^{B^{N}}\cdots\Pi_{\left(  N\right)  ,u_{1}^{N-1}u_{N}%
}^{B^{N}}\bigg\},
\end{align*}
where we are assuming a particular choice of the bits $u_{\mathcal{A}^{c}}$ in
the sequence of projectors $\Pi_{\left(  N\right)  ,u_{1}^{N-1}u_{N}}^{B^{N}}$
$\cdots$ $\Pi_{\left(  i\right)  ,u_{1}^{i-1}u_{i}}^{B^{N}}$ $\cdots$
$\Pi_{\left(  1\right)  ,u_{1}}^{B^{N}}$ and the convention mentioned before
that $\Pi_{\left(  i\right)  ,u_{1}^{i-1}u_{i}}^{B^{N}}=I$ if $u_{i}$ is a
frozen bit. We are also assuming that the sender transmits the information
sequence $u_{\mathcal{A}}$ with uniform probability $2^{-K}$. The probability
of error $P_{e}\left(  N,K,\mathcal{A}\right)  $ averaged over all choices of
the frozen bits is then%
\begin{align}
&  P_{e}\left(  N,K,\mathcal{A}\right)  \nonumber\\
&  =\frac{1}{2^{N-K}}\sum_{u_{\mathcal{A}^{c}}}P_{e}\left(  N,K,\mathcal{A}%
,u_{\mathcal{A}^{c}}\right)  \nonumber\\
&  =1-\frac{1}{2^{N}}\sum_{u^{N}}\text{Tr}\bigg\{\Pi_{\left(  N\right)
,u_{1}^{N-1}u_{N}}^{B^{N}}\cdots\Pi_{\left(  i\right)  ,u_{1}^{i-1}u_{i}%
}^{B^{N}}\cdots\nonumber\\
&  \ \ \ \ \cdots\Pi_{\left(  1\right)  ,u_{1}}^{B^{N}}\ \rho_{u^{N}}%
\ \Pi_{\left(  1\right)  ,u_{1}}^{B^{N}}\cdots\Pi_{\left(  i\right)
,u_{1}^{i-1}u_{i}}^{B^{N}}\cdots\Pi_{\left(  N\right)  ,u_{1}^{N-1}u_{N}%
}^{B^{N}}\bigg\}.\label{eq:avg-error-prob}%
\end{align}

One of the main contributions of this paper is the following proposition regarding the
average ensemble performance of polar codes with a quantum successive
cancellation decoder:

\begin{proposition}
\label{prop:error-bound}For any classical-quantum channel $W$ with binary
inputs and quantum outputs and any choice of $\left(  N,K,\mathcal{A}\right)
$, the following bound holds%
\[
P_{e}\left(  N,K,\mathcal{A}\right)  \leq2\sqrt{\sum_{i\in\mathcal{A}}\frac
{1}{2}\sqrt{F(W_{N}^{\left(  i\right)  })}}.
\]
Thus, there exists a frozen vector $u_{\mathcal{A}^{c}}$ for each $\left(
N,K,\mathcal{A}\right)  $ such that%
\[
P_{e}\left(  N,K,\mathcal{A},u_{\mathcal{A}^{c}}\right)  \leq2\sqrt{\sum
_{i\in\mathcal{A}}\frac{1}{2}\sqrt{F(W_{N}^{\left(  i\right)  })}}.
\]

\end{proposition}

\subsubsection{Polar Coding Theorem}

Proposition~\ref{prop:error-bound} immediately leads to the definition of
polar codes for classical-quantum channels:

\begin{definition}
[Polar Code]\label{def:polar-code}A polar code for $W$ is a $G_{N}$-coset code
with parameters $\left(  N,K,\mathcal{A},u_{\mathcal{A}^{c}}\right)  $ where
the information set $\mathcal{A}$ is such that $\left\vert \mathcal{A}%
\right\vert =K$ and%
\[
F(W_{N}^{\left(  i\right)  })\leq F(W_{N}^{\left(  j\right)  })\text{ for all
}i\in\mathcal{A}\text{ and }j\in\mathcal{A}^{c}.
\]

\end{definition}

We can finally state the polar coding theorem for classical-quantum channels.
Consider a classical-quantum channel $W$ and a real number $R\geq0$. Let%
\[
P_{e}\left(  N,R\right)  =P_{e}\left(  N,\left\lfloor NR\right\rfloor
,\mathcal{A}\right)  ,
\]
with the information bit set chosen according to the polar coding rule in
Definition~\ref{def:polar-code}. So $P_{e}\left(  N,\mathcal{A}\right)  $ is
the block error probability for polar coding over $W$ with blocklength $N$,
rate $R$, and quantum successive cancellation decoding averaged uniformly over
the frozen bits $u_{\mathcal{A}^{c}}$.

\begin{theorem}
[Polar Coding Theorem]\label{thm:polar-coding-theorem}For any
classical-quantum channel $W$ with binary inputs and quantum outputs, a fixed
$R<I\left(  W\right)  $, and $\beta<1/2$, the block error probability
$P_{e}\left(  N,R\right)  $ satisfies the following bound:%
\[
P_{e}\left(  N,R\right)  =o(2^{-\frac{1}{2}N^{\beta}}).
\]

\end{theorem}

The polar coding theorem above follows as a straightforward corollary of
Theorem~\ref{thm:polar-rate} and Proposition~\ref{prop:error-bound}.

\section{Recursive Channel Transformations}

\label{sec:rec-chan-trans}This section delves into more detail regarding
recursive channel combining and channel splitting. Recall the channel
combining in (\ref{eq:channel-combine-1}-\ref{eq:GN-matrix}) and the channel
splitting in (\ref{eq:split-channels}). These allowed for us to take $N$
independent copies of a classical-quantum channel $W^{\otimes N}$ and
transform them into the $N$ split channels $W_{N}^{\left(  1\right)  }$,
\ldots, $W_{N}^{\left(  N\right)  }$. We show here how to break the channel
transformation into a series of single-step transformations. Much of the
discussion here parallels Arikan's discussion in Sections~II and III of
Ref.~\cite{A09}.

\begin{figure}[ptb]
\begin{center}
\includegraphics[
width=2.3263in
]{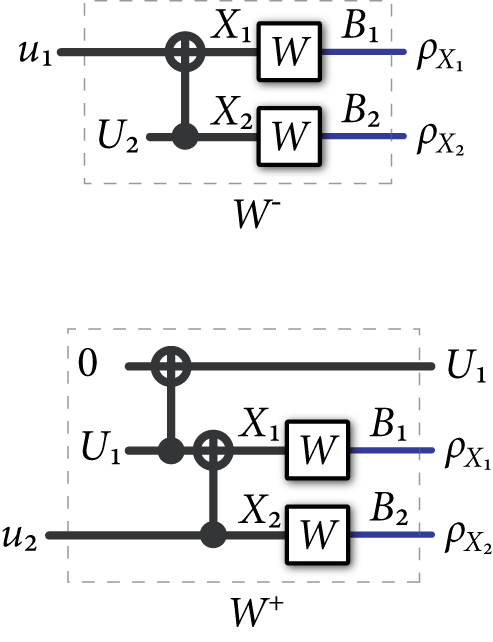}
\end{center}
\caption{The channels $W^{-}$ and $W^{+}$ induced from channel combining and
channel splitting. The channel $W^{-}$ with input $u_1$ is induced by selecting the bit $U_2$ uniformly
at random, passing both $u_1$ and $U_2$ through the encoder, and then through the two channel uses.
The channel $W^{+}$ with input $u_2$ is induced by selecting $U_1$ uniformly at random,
copying it to another bit (via the classical CNOT gate), sending both $U_1$ and $u_2$
through the encoder, and the outputs are the quantum outputs and the bit $U_1$. }%
\label{fig:split-channels}%
\end{figure}

We obtain a pair of channels $W^{-}$ and $W^{+}$ from two independent copies
of a channel $W:x\rightarrow\rho_{x}$ by a single-step transformation if it
holds that%
\[
W^{-}:u_{1}\rightarrow\rho_{u_{1}}^{-},
\]
where%
\begin{equation}
\rho_{u_{1}}^{-}\equiv\sum_{u_{2}}\frac{1}{2}\rho_{u_{2}\oplus u_{1}}^{B_{1}%
}\otimes\rho_{u_{2}}^{B_{2}}. \label{eq:channel-minus}%
\end{equation}
Also, it should hold that%
\[
W^{+}:u_{2}\rightarrow\rho_{u_{2}}^{+},
\]
where%
\begin{align}
\rho_{u_{2}}^{+}  &  \equiv\sum_{u_{1}}\frac{1}{2}\left\vert u_{1}%
\right\rangle \left\langle u_{1}\right\vert ^{U_{1}}\otimes\rho_{u_{2}\oplus
u_{1}}^{B_{1}}\otimes\rho_{u_{2}}^{B_{2}}\label{eq:channel-plus}\\
&  =\left(  \sum_{u_{1}}\frac{1}{2}\left\vert u_{1}\right\rangle \left\langle
u_{1}\right\vert ^{U_{1}}\otimes\rho_{u_{2}\oplus u_{1}}^{B_{1}}\right)
\otimes\rho_{u_{2}}^{B_{2}}.\nonumber
\end{align}
We use the following notation to denote such a
transformation:%
\[
\left(  W,W\right)  \rightarrow\left(  W^{-},W^{+}\right)  .
\]
Additionally, we choose the notation $W^{-}$ and $W^{+}$ so that $W^{-}$
denotes the worse channel and $W^{+}$ denotes the better channel.
Figure~\ref{fig:split-channels}\ depicts the channels $W^{-}$ and $W^{+}%
$.

Thus, from the above, we can write $\left(  W,W\right)  \rightarrow
(W_{2}^{\left(  1\right)  },W_{2}^{\left(  2\right)  })$ because, by the
definition in (\ref{eq:split-channels}), we have%
\begin{align*}
W_{2}^{\left(  1\right)  }\left(  u_{1}\right)   &  =\sum_{u_{2}}\frac{1}%
{2}\rho_{u_{1}\oplus u_{2}}^{B_{1}}\otimes\rho_{u_{2}}^{B_{2}},\\
W_{2}^{\left(  2\right)  }\left(  u_{2}\right)   &  =\sum_{u_{1}}\frac{1}%
{2}\left\vert u_{1}\right\rangle \left\langle u_{1}\right\vert ^{U_{1}}%
\otimes\rho_{u_{1}\oplus u_{2}}^{B_{1}}\otimes\rho_{u_{2}}^{B_{2}}.
\end{align*}
We can actually write more generally%
\begin{equation}
(W_{N}^{\left(  i\right)  },W_{N}^{\left(  i\right)  })\rightarrow
(W_{2N}^{\left(  2i-1\right)  },W_{2N}^{\left(  2i\right)  }),
\label{eq:general-rec-trans}%
\end{equation}
which follows as a corollary to

\begin{proposition}
\label{prop:rec-combos}For any $n\geq0$, $N=2^{n}$, and $1\leq i\leq N$, it
holds that%
\begin{align}
W_{2N}^{\left(  2i-1\right)  }\left(  u_{2i-1}\right)   &  =\sum_{u_{2i}}%
\frac{1}{2}W_{N}^{\left(  i\right)  }\left(  u_{2i-1}\oplus u_{2i}\right)
\otimes W_{N}^{\left(  i\right)  }\left(  u_{2i}\right)
,\label{eq:recursive-trans-1}\\
W_{2N}^{\left(  2i\right)  }\left(  u_{2i}\right)   &  =W_{N}^{\left(
i\right)  }\left(  u_{2i-1}\oplus u_{2i}\right)  \otimes W_{N}^{\left(
i\right)  }\left(  u_{2i}\right)  , \label{eq:recursive-trans-2}%
\end{align}
with $W_{N}^{\left(  i\right)  }$ defined in (\ref{eq:split-channels}).
\end{proposition}

\begin{IEEEproof}
The proof of the above proposition is similar to the proof of Arikan's
Proposition~3 \cite{A09}.
\end{IEEEproof}

We can justify the relationship in (\ref{eq:general-rec-trans}) by observing
that (\ref{eq:recursive-trans-1}) and (\ref{eq:recursive-trans-2}) have the
same form as (\ref{eq:channel-minus}) and (\ref{eq:channel-plus}) with the
following substitutions:%
\begin{align*}
W  &  \leftarrow W_{N}^{\left(  i\right)  },\\
W^{+}  &  \leftarrow W_{2N}^{\left(  2i\right)  },\\
W^{-}  &  \leftarrow W_{2N}^{\left(  2i-1\right)  },\\
u_{1}  &  \leftarrow u_{2i-1},\\
u_{2}  &  \leftarrow u_{2i}.
\end{align*}

\subsection{Transformation of Rate and Reliability}

This section considers how both the rate $I(W_{N}^{(i)})$ and reliability
$F(W_{N}^{(i)})$ evolve under the general transformation in
(\ref{eq:general-rec-trans}). All proofs of the results in this section appear
in the appendix.

\begin{proposition}
\label{prop:rate-cons-polarize}Suppose that $(W,W)\rightarrow(W^{-},W^{+})$
for some channels satisfying (\ref{eq:channel-minus}-\ref{eq:channel-plus}).
Then the following rate conservation and polarizing relations hold%
\begin{align}
I\left(  W^{-}\right)  +I\left(  W^{+}\right)   &  =2I\left(  W\right)  ,\\
I\left(  W^{-}\right)   &  \leq I\left(  W^{+}\right)  .
\end{align}

\end{proposition}

We can conclude from the above two relations that%
\[
I\left(  W^{-}\right)  \leq I\left(  W\right)  \leq I\left(  W^{+}\right)  .
\]

The following proposition states how the reliability evolves under the channel transformation:

\begin{proposition}
\label{prop:rel-trans}Suppose $\left(  W,W\right)  \rightarrow\left(
W^{-},W^{+}\right)  $ for some channels satisfying (\ref{eq:channel-minus}%
-\ref{eq:channel-plus}). Then%
\begin{align}
\sqrt{F\left(  W^{+}\right)  }  &  =F\left(  W\right)
,\label{eq:fidelity-trans-1}\\
\sqrt{F\left(  W^{-}\right)  }  &  \leq2\sqrt{F\left(  W\right)  }-F\left(
W\right)  ,\label{eq:fidelity-trans-2}\\
F\left(  W^{-}\right)   &  \geq F\left(  W\right)  \geq F\left(  W^{+}\right)
. \label{eq:fidelity-trans-3}%
\end{align}

\end{proposition}

By combining (\ref{eq:fidelity-trans-1}) with (\ref{eq:fidelity-trans-2}), we
observe that the reliability only improves under a single-step transformation:%
\[
\sqrt{F\left(  W^{-}\right)  }+\sqrt{F\left(  W^{+}\right)  }\leq
2\sqrt{F\left(  W\right)  }.
\]

The above propositions for the single-step transformation lead us to the
following proposition in the general case:

\begin{proposition}
\label{prop:rec-rate-rel}For any classical-quantum channel $W$, $N=2^{n}$,
$n\geq0$, and $1\leq i\leq N$, the local transformation in
(\ref{eq:general-rec-trans}) preserves rate and improves reliability in the
following sense:%
\begin{align}
I(W_{2N}^{\left(  2i-1\right)  })+I(W_{2N}^{\left(  2i\right)  })  &
=2I(W_{N}^{\left(  i\right)  }),\label{eq:single-rate}\\
\sqrt{F(W_{2N}^{\left(  2i-1\right)  })}+\sqrt{F(W_{2N}^{\left(  2i\right)
})}  &  \leq2\sqrt{F(W_{N}^{\left(  i\right)  })}. \label{eq:single-rel}%
\end{align}
Channel splitting moves rate and reliability \textquotedblleft away from the
center\textquotedblright:%
\begin{align*}
I(W_{2N}^{\left(  2i-1\right)  })  &  \leq I(W_{N}^{\left(  i\right)  })\leq
I(W_{2N}^{\left(  2i\right)  }),\\
\sqrt{F(W_{2N}^{\left(  2i-1\right)  })}  &  \geq\sqrt{F(W_{N}^{\left(
i\right)  })}\geq\sqrt{F(W_{2N}^{\left(  2i\right)  })}.
\end{align*}
The reliability terms satisfy%
\begin{align}
\sqrt{F(W_{2N}^{\left(  2i\right)  })}  &  =F(W_{N}^{\left(  i\right)  }),\\
\sqrt{F(W_{2N}^{\left(  2i-1\right)  })}  &  \leq2\sqrt{F(W_{N}^{\left(
i\right)  })}-F(W_{N}^{\left(  i\right)  }),
\end{align}
and the cumulative rate and reliability satisfy%
\begin{align}
\sum_{i=1}^{N}I(W_{N}^{\left(  i\right)  })  &  =N\ I\left(  W\right)
,\label{eq:cumul-rate}\\
\sum_{i=1}^{N}\sqrt{F(W_{N}^{\left(  i\right)  })}  &  \leq N\ \sqrt{F\left(
W\right)  }. \label{eq:cumul-rel}%
\end{align}

\end{proposition}

The above proposition follows directly from Propositions~\ref{prop:rec-combos}%
, \ref{prop:rate-cons-polarize}, and \ref{prop:rel-trans}. The relations in
(\ref{eq:cumul-rate}) and (\ref{eq:cumul-rel}) follow from applying
(\ref{eq:single-rate}) and (\ref{eq:single-rel}) repeatedly.

\section{Channel Polarization}

\label{sec:channel-polarization}We are now in a position to prove
Theorem~\ref{thm:channel-polarization}\ on channel polarization. The idea
behind the proof of this theorem is identical to Arikan's proof of his
Theorem~1 in Ref.~\cite{A09}---with the relationships in
Propositions~\ref{prop:rate-cons-polarize} and \ref{prop:rel-trans} already
established, we can readily exploit the martingale proof technique. Thus, we
only provide a brief summary of the proof of
Theorem~\ref{thm:channel-polarization} by following the presentation in
Chapter~2 of Ref.~\cite{K09}.

Consider the channel $W_{N}^{\left(  i\right)  }$. Let $b_{1}\cdots b_{n}$
denote an $n$-bit binary expansion of the channel index $i$ and let
$W_{\left(  b_{1}\cdots b_{n}\right)  }=W_{N}^{\left(  i\right)  }$. Then we
can construct the channel $W_{\left(  b_{1}\cdots b_{k}\right)  }$ by
combining two copies of $W_{\left(  b_{1}\cdots b_{k-1}\right)  }$ according
to (\ref{eq:recursive-trans-1}) if $b_{k}=0$ or by combining two copies of
$W_{\left(  b_{1}\cdots b_{k-1}\right)  }$ according to
(\ref{eq:recursive-trans-2}) if $b_{k}=1$. We repeatedly construct all the way
from $b_{1}$ until $b_{n}$ with the above rule.

Arikan's idea was to represent the channel construction as a random birth
process in order to analyze its limiting behavior. In order to do so, we let
$\left\{  B_{n}:n\geq1\right\}  $ be a sequence of IID\ uniform Bernoulli
random variables, where we define each over a probability space $(\Omega
,\mathcal{F},P)$. Let $\mathcal{F}_{0}$ denote the trivial $\sigma$-field.
Also, let $\left\{  \mathcal{F}_{n}:n\geq1\right\}  $ denote the $\sigma
$-fields that the random variables $\left(  B_{1},\ldots,B_{n}\right)  $
generate. We also assume that $\mathcal{F}_{0}\subseteq\mathcal{F}%
_{1}\subseteq\mathcal{\cdots}\subseteq\mathcal{F}_{n}$. Let $W_{0}=W$ and let
$\left\{  W_{n}:n\geq0\right\}  $ denote a sequence of operator-valued random
variables that forms a tree process where $W_{n+1}$ is constructed from two
copies of $W_{n}$ according to (\ref{eq:recursive-trans-1}) if $B_{n}=0$ and
according to (\ref{eq:recursive-trans-2}) if $B_{n}=1$. The output space of
the operator-valued random variable $W_{n}$ is equal to $\{W_{2^{n}}^{\left(
i\right)  }\}_{i=1}^{2^{n}}$. We are not really concerned with the channel
process $\left\{  W_{n}:n\geq0\right\}  $ but more so with the fidelities
$\{F(W_{N}^{\left(  i\right)  })\}$ and Holevo informations $\{I(W_{N}%
^{\left(  i\right)  })\}$. Thus, we can simply analyze the limiting behavior
of the two random processes $\{F_{n}:n\geq0\}\equiv\{\sqrt{F(W_{n})}:n\geq0\}$
and $\{I_{n}:n\geq0\}\equiv\{I(W_{n}):n\geq0\}$. By the definitions of the
random variables $F_{n}$ and $I_{n}$, it follows that%
\begin{align*}
\Pr\left\{  I_{n}\in\left(  a,b\right)  \right\}   &  =\frac{1}{2^{n}%
}|\{i:I(W_{2^{n}}^{\left(  i\right)  })\in\left(  a,b\right)  \}|,\\
\Pr\left\{  F_{n}\in\left(  a,b\right)  \right\}   &  =\frac{1}{2^{n}%
}|\{i:F(W_{2^{n}}^{\left(  i\right)  })\in\left(  a,b\right)  \}|.
\end{align*}
We then have the following lemma.

\begin{lemma}
The sequence $\{\left(  F_{n},\mathcal{F}_{n}\right)  :n\geq0\}$ is a bounded
super-martingale, and the sequence $\{\left(  I_{n},\mathcal{F}_{n}\right)
:n\geq0\}$ is a bounded martingale.
\end{lemma}

\begin{IEEEproof}
Let $b_{1}\cdots b_{n}$ be a particular realization of the random sequence
$B_{1}\cdots B_{n}$. Then the conditional expectation satisfies%
\begin{align*}
&  \mathbb{E}\left\{  I_{n+1}\ |\ B_{1}=b_{1},\ldots,B_{n}=b_{n}\right\}  \\
&  =\frac{1}{2}I\left(  W_{\left(  b_{1},\ldots,b_{n},0\right)  }\right)
+\frac{1}{2}I\left(  W_{\left(  b_{1},\ldots,b_{n},1\right)  }\right)  \\
&  =I\left(  W_{\left(  b_{1},\ldots,b_{n}\right)  }\right)  \\
&  =I_{n},
\end{align*}
where the second equality follows from the definition of $W_{\left(
b_{1},\ldots,b_{n},0\right)  }$ and $W_{\left(  b_{1},\ldots,b_{n},1\right)
}$ and Proposition~\ref{prop:rec-rate-rel}. The proof for $\left\{
F_{n}\right\}  $ similarly follows from the definitions and
Proposition~\ref{prop:rec-rate-rel}. The boundedness condition follows because
$0\leq I\left(  W\right)  ,F\left(  W\right)  \leq1$ for any classical-quantum
channel $W$ with binary inputs and quantum outputs.
\end{IEEEproof}

We can now finally prove Theorem~\ref{thm:channel-polarization}\ regarding
channel polarization. Given that $\left\{  I_{n}\right\}  $ is a bounded
martingale and $\left\{  F_{n}\right\}  $ is a bounded super-martingale, the
limits $\lim_{n\rightarrow\infty}I_{n}$ and $\lim_{n\rightarrow\infty}F_{n}$
converge almost surely and in $\mathcal{L}_{1}$ to the random variables
$I_{\infty}$ and $F_{\infty}$. The convergence implies that $\mathbb{E}%
\left\{  \left\vert F_{n+1}-F_{n}\right\vert \right\}  \rightarrow0$ as
$n\rightarrow\infty$. By the definition of the process $\left\{
F_{n}\right\}  $, it holds that $F_{n+1}=F_{n}^{2}$ with probability $\frac
{1}{2}$, so that%
\[
\mathbb{E}\left\{  \left\vert F_{n+1}-F_{n}\right\vert \right\}  \geq\frac
{1}{2}\mathbb{E}\left\{  \left\vert F_{n}\left(  1-F_{n}\right)  \right\vert
\right\}  \geq0.
\]
It then follows that $\mathbb{E}\left\{  \left\vert F_{n}\left(
1-F_{n}\right)  \right\vert \right\}  \rightarrow0$ as $n\rightarrow\infty$,
which in turn implies that $\mathbb{E}\left\{  \left\vert F_{\infty}\left(
1-F_{\infty}\right)  \right\vert \right\}  =0$. We conclude that $F_{\infty
}\in\left\{  0,1\right\}  $ almost surely. Combining this result with
Proposition~\ref{prop:IvsF} proves that $I_{\infty}\in\left\{  0,1\right\}  $
almost surely. Finally, we have that $\Pr\left\{  I_{\infty}=1\right\}
=\mathbb{E}\left\{  I_{\infty}\right\}  =\mathbb{E}\left\{  I_{0}\right\}
=I\left(  W\right)  $ because $I_{n}$ is a martingale.

\section{Performance of Polar Coding}

\label{sec:succ-canc-analysis}We can now analyze the performance under the
above successive cancellation decoding scheme and provide a proof of
Proposition~\ref{prop:error-bound}. The proof of
Theorem~\ref{thm:polar-coding-theorem} readily follows by applying
Proposition~\ref{prop:error-bound} and Theorem~\ref{thm:polar-rate}.

First recall the following \textquotedblleft non-commutative union
bound\textquotedblright\ of Sen (Lemma~3 in Ref.~\cite{S11}):%
\begin{equation}
1-\text{Tr}\left\{  \Pi_{N}\cdots\Pi_{1}\rho\Pi_{1}\cdots\Pi_{N}\right\}
\leq2\sqrt{\sum_{i=1}^{N}\text{Tr}\left\{  \left(  I-\Pi_{i}\right)
\rho\right\}  },\label{eq:sen-bound}%
\end{equation}
which holds for projectors $\Pi_{1}$, \ldots, $\Pi_{N}$ and a density operator
$\rho$.\footnote{We say that Sen's bound is a \textquotedblleft
non-commutative union bound\textquotedblright\ because it is analogous to the
following union bound from probability theory: $\Pr\left\{  \left(  A_{1}%
\cap\cdots\cap A_{N}\right)  ^{c}\right\}  =\Pr\left\{  A_{1}^{c}\cup
\cdots\cup A_{N}^{c}\right\}  \leq\sum_{i=1}^{N}\Pr\left\{  A_{i}^{c}\right\}
$, where $A_{1}$, \ldots, $A_{N}$ are events. The analogous bound for
projector logic would be Tr$\left\{  \left(  I-\Pi_{1}\cdots\Pi_{N}\cdots
\Pi_{1}\right)  \rho\right\}  \leq\sum_{i=1}^{N}$Tr$\left\{  \left(  I-\Pi
_{i}\right)  \rho\right\}  $, if we think of $\Pi_{1}\cdots\Pi_{N}$ as a
projector onto the intersection of subspaces. Though, the above bound only
holds if the projectors $\Pi_{1}$, \ldots, $\Pi_{N}$ are commuting (choosing
$\Pi_{1}=\left\vert +\right\rangle \left\langle +\right\vert $, $\Pi
_{2}=\left\vert 0\right\rangle \left\langle 0\right\vert $, and $\rho
=\left\vert 0\right\rangle \left\langle 0\right\vert $ gives a
counterexample). If the projectors are non-commuting, then Sen's bound in
(\ref{eq:sen-bound}) is the next best thing and suffices for our purposes
here.} We begin by applying the above inequality to $P_{e}\left(
N,K,\mathcal{A}\right)  $ (defined in (\ref{eq:avg-error-prob})):%
\begin{align*}
&  P_{e}\left(  N,K,\mathcal{A}\right)  \\
&  =\frac{1}{2^{N}}\sum_{u^{N}}\bigg(1-\text{Tr}\bigg\{\Pi_{\left(  N\right)
,u_{1}^{N-1}u_{N}}^{B^{N}}\cdots\Pi_{\left(  i\right)  ,u_{1}^{i-1}u_{i}%
}^{B^{N}}\cdots\\
&  \ \ \ \ \cdots\Pi_{\left(  1\right)  ,u_{1}}^{B^{N}}\ \rho_{u^{N}}%
\ \Pi_{\left(  1\right)  ,u_{1}}^{B^{N}}\cdots\Pi_{\left(  i\right)
,u_{1}^{i-1}u_{i}}^{B^{N}}\cdots\Pi_{\left(  N\right)  ,u_{1}^{N-1}u_{N}%
}^{B^{N}}\bigg\}\bigg)\\
&  \leq\frac{1}{2^{N}}\sum_{u^{N}}2\sqrt{\sum_{i=1}^{N}\text{Tr}\left\{
\left(  I-\Pi_{\left(  i\right)  ,u_{1}^{i-1}u_{i}}^{B^{N}}\right)
\rho_{u^{N}}\right\}  }\\
&  =\frac{1}{2^{N}}\sum_{u^{N}}2\sqrt{\sum_{i\in\mathcal{A}}\text{Tr}\left\{
\left(  I-\Pi_{\left(  i\right)  ,u_{1}^{i-1}u_{i}}^{B^{N}}\right)
\rho_{u^{N}}\right\}  }\\
&  \leq2\sqrt{\frac{1}{2^{N}}\sum_{u^{N}}\sum_{i\in\mathcal{A}}\text{Tr}%
\left\{  \left(  I-\Pi_{\left(  i\right)  ,u_{1}^{i-1}u_{i}}^{B^{N}}\right)
\rho_{u^{N}}\right\}  }%
\end{align*}
where the second equality follows from our convention that $\Pi_{\left(
i\right)  ,u_{1}^{i-1}u_{i}}^{B^{N}}=I$ if $u_{i}$ is a frozen bit and the
second inequality follows from concavity of the square root. Continuing, we
have%
\begin{align*}
&  =2\sqrt{\sum_{i\in\mathcal{A}}\sum_{u^{N}}\frac{1}{2^{N}}\text{Tr}\left\{
\hat{\Pi}_{\left(  i\right)  ,u_{1}^{i-1}u_{i}}^{B^{N}}\rho_{u^{N}}\right\}
}\\
&  =2\sqrt{\sum_{i\in\mathcal{A}}\sum_{u_{1}^{i-1}}\frac{1}{2^{i-1}}%
\sum_{u_{i}}\frac{1}{2}\sum_{u_{i+1}^{N}}\frac{1}{2^{N-i}}\text{Tr}\left\{
\hat{\Pi}_{\left(  i\right)  ,u_{1}^{i-1}u_{i}}^{B^{N}}\rho_{u^{N}}\right\}
}\\
&  =2\sqrt{\sum_{i\in\mathcal{A}}\sum_{u_{1}^{i-1}}\frac{1}{2^{i-1}}%
\sum_{u_{i}}\frac{1}{2}\text{Tr}\left\{  \hat{\Pi}_{\left(  i\right)
,u_{1}^{i-1}u_{i}}^{B^{N}}\sum_{u_{i+1}^{N}}\frac{1}{2^{N-i}}\rho_{u^{N}%
}\right\}  },
\end{align*}
where we define%
\[
\hat{\Pi}_{\left(  i\right)  ,u_{1}^{i-1}u_{i}}^{B^{N}}=I-\Pi_{\left(
i\right)  ,u_{1}^{i-1}u_{i}}^{B^{N}}.
\]
The first equality follows from exchanging the sums. The second equality
follows from expanding the sum and normalization $\sum_{u^{N}}\frac{1}{2^{N}}%
$. The third equality follows from bringing the sum $\sum_{u_{i+1}^{N}}%
\frac{1}{2^{N-i}}$ inside the trace. Continuing,%
\begin{align*}
&  =2\sqrt{\sum_{i\in\mathcal{A}}\sum_{u_{1}^{i-1}}\frac{1}{2^{i-1}}%
\sum_{u_{i}}\frac{1}{2}\text{Tr}\left\{  \left(  I-\Pi_{\left(  i\right)
,u_{1}^{i-1}u_{i}}^{B^{N}}\right)  \overline{\rho}_{u_{1}^{i}}^{B^{N}%
}\right\}  }\\
&  =2\sqrt{\sum_{i\in\mathcal{A}}\sum_{u_{i}}\frac{1}{2}\sum_{u_{1}^{i-1}%
}\frac{1}{2^{i-1}}\text{Tr}\left\{  \left(  I-\Pi_{\left(  i\right)
,u_{1}^{i-1}u_{i}}^{B^{N}}\right)  \overline{\rho}_{u_{1}^{i}}^{B^{N}%
}\right\}  }\\
&  =2\bigg(\sum_{i\in\mathcal{A}}\sum_{u_{i}}\frac{1}{2}\text{Tr}\big\{(I-\sum
_{u_{1}^{i-1}}\left\vert u_{1}^{i-1}\right\rangle \left\langle u_{1}%
^{i-1}\right\vert ^{U_{1}^{i-1}}\otimes\Pi_{\left(  i\right)  ,u_{1}%
^{i-1}u_{i}}^{B^{N}})\\
&  \ \ \ \ \ \ \ \ \ \ \ \ \sum_{u_{1}^{i-1}}\frac{1}{2^{i-1}}\left\vert
u_{1}^{i-1}\right\rangle \left\langle u_{1}^{i-1}\right\vert ^{U_{1}^{i-1}%
}\otimes\overline{\rho}_{u_{1}^{i}}^{B^{N}}\big\}\bigg)^{-\frac{1}{2}}%
\end{align*}
The first equality is from the definition in (\ref{eq:averaged-cond-states}).
The second equality is from exchanging sums. The third equality is from the
fact that%
\begin{multline*}
\sum_{x}p\left(  x\right)  \text{Tr}\left\{  A_{x}\rho_{x}\right\}  =\\
\text{Tr}\left\{  \left(  \sum_{x}\left\vert x\right\rangle \left\langle
x\right\vert \otimes A_{x}\right)  \left(  \sum_{x^{\prime}}p\left(
x^{\prime}\right)  \left\vert x^{\prime}\right\rangle \left\langle x^{\prime
}\right\vert \otimes\rho_{x^{\prime}}\right)  \right\}  .
\end{multline*}
Continuing,%
\begin{align*}
&  =2\sqrt{\sum_{i\in\mathcal{A}}\sum_{u_{i}}\frac{1}{2}\text{Tr}\left\{
\left(  I-\Pi_{\left(  i\right)  ,u_{i}}^{U_{1}^{i-1}B^{N}}\right)
\rho_{\left(  i\right)  ,u_{i}}^{U_{1}^{i-1}B^{N}}\right\}  }\\
&  \leq2\sqrt{\sum_{i\in\mathcal{A}}\frac{1}{2}\sqrt{F\left(  W^{\left(
i\right)  }\right)  }}%
\end{align*}
The first equality is from the observations in (\ref{eq:projectors-expanded-1}%
-\ref{eq:projectors-expanded-2}) and the definition in
(\ref{eq:split-channels}). The final inequality follows from Lemma 3.2 of
Ref.~\cite{H06} and the definition in (\ref{eq:channel-fidelity}). This
completes the proof of Proposition~\ref{prop:error-bound}.

We state the proof of Theorem~\ref{thm:polar-coding-theorem}\ for
completeness. Invoking Theorem~\ref{thm:polar-rate}, there exists a sequence
of sets $\mathcal{A}_{N}$ with size $\left\vert \mathcal{A}_{N}\right\vert
\geq NR$ for any $R<I\left(  W\right)  $ and $\beta<1/2$ such that%
\[
\sum_{i\in\mathcal{A}_{N}}\sqrt{F\left(  W^{\left(  i\right)  }\right)
}=o(2^{-N^{\beta}}),
\]
and thus%
\[
2\sqrt{\sum_{i\in\mathcal{A}_{N}}\frac{1}{2}\sqrt{F\left(  W^{\left(
i\right)  }\right)  }}=o(2^{-\frac{1}{2}N^{\beta}}).
\]
This bound holds if we choose the set $\mathcal{A}_{N}$ according to the polar
coding rule because this rule minimizes the above sum by definition.
Theorem~\ref{thm:polar-coding-theorem}\ follows by combining
Proposition~\ref{prop:error-bound}\ with this fact about the polar coding rule.

\section{Conclusion}

\label{sec:concl}

We have shown how to construct polar codes for channels with classical binary
inputs and quantum outputs, and we showed that they can achieve the symmetric Holevo
information rate for classical communication.
In fact, for a quantum channel with binary pure state outputs,
such as a binary-phase-shift-keyed (BPSK) coherent-state optical communication
alphabet, the symmetric Holevo information rate is the ultimate
channel capacity \cite{G11}, which is therefore achieved by our polar code \cite{GW12}.
The general idea behind the construction is
similar to Arikan's \cite{A09}, but we required several technical advances in
order to demonstrate both channel polarization at the symmetric
Holevo information rate and the operation of the
quantum successive cancellation decoder. To prove that channel polarization
takes hold, we could exploit several results in the quantum information
literature \cite{LR73,FvG99,H00,NC00,RFZ10,W11} and some of Arikan's tools. To
prove that the quantum successive cancellation decoder works well, we
exploited some ideas from quantum hypothesis testing
\cite{H69,Hol72,Hel76,H06,NH07} and Sen's recent ``non-commutative union
bound'' \cite{S11}. The result is a near-explicit code construction that
achieves the symmetric Holevo information rate for channels with classical inputs and quantum outputs.
(When we say ``near-explicit,'' we mean that it still remains open in the quantum case to
determine which synthesized channels are good or bad.) 
Also, several works have now appeared on polar coding for private classical communication
and quantum communication \cite{WG11a,RDR11,WG11a,WR12,WR12a,DGW12},
most of which use the results developed in this paper.

One of the main open problems going forward from here is to simplify the quantum
successive cancellation decoder. Arikan could show how to calculate later
estimates by exploiting the results of earlier estimates in an ``FFT-like''
fashion, and this observation reduced the complexity of the decoding to $O(N
\log N)$. It is not clear to us yet how to reduce the complexity of the
quantum successive cancellation decoder because it is not merely a matter of
computing formulas, but rather a sequence of physical operations
(measurements) that the receiver needs to perform on the channel output
systems. If there were some way to perform the measurements on smaller systems
and then adaptively perform other measurements based on earlier results, then
this would be helpful in demonstrating a reduced complexity.

Another important open question is to devise an efficient
construction of the polar codes, something that remains an
open problem even for classical polar codes. However, there has been recent
work on efficient suboptimal classical polar code constructions \cite{TV11},
which one might try to extend to polar codes for the
classical-quantum channel. Finally, extending our code and decoder construction
to a classical-quantum channel with a non-binary (M-ary) alphabet
remains a good open line for investigation.

\section{Acknowledgments}

MMW acknowledges financial support from the MDEIE (Qu\'{e}bec) PSR-SIIRI
international collaboration grant. SG was supported by the DARPA Information
in a Photon (InPho) program under contract number HR0011-10-C-0159. We thank
David Forney, MIT for suggesting us to try polar codes for the quantum
channel. We also thank Emre Telatar, EPFL for an intuitive tutorial on channel
polarization at ISIT 2011.

\appendix


\begin{IEEEproof}
[Proof of Proposition~\ref{prop:IvsF}]The first bound in
(\ref{eq:rate-rel-1st-bound}) follows from Holevo's characterization of the
quantum cutoff rate (Proposition~1 of Ref.~\cite{H00}). In particular, Holevo
proved that the following inequality holds for all $s\in\left[  0,1\right]  $:%
\[
I\left(  X;B\right)  _{\omega}\geq-\log\text{Tr}\left\{  \left(  \sum
_{x\in\mathcal{X}}p_{X}\left(  x\right)  \left(  \omega_{x}\right)  ^{\frac
{1}{1+s}}\right)  ^{1+s}\right\}  ,
\]
where the entropy on the LHS\ is with respect to a classical-quantum state%
\[
\omega^{XB}\equiv\sum_{x\in\mathcal{X}}p_{X}\left(  x\right)  \left\vert
x\right\rangle \left\langle x\right\vert ^{X}\otimes\omega_{x}^{B}.
\]
By setting $s=1$, the alphabet $\mathcal{X}=\left\{  0,1\right\}  $, and the
distribution $p_{X}\left(  x\right)  $ to be uniform, we obtain the bound%
\begin{align*}
I\left(  W\right)   &  \geq-\log\left(  \text{Tr}\left\{  \left(  \frac{1}%
{2}\left(  \sqrt{\rho_{0}}+\sqrt{\rho_{1}}\right)  \right)  ^{2}\right\}
\right) \\
&  =-\log\left(  \frac{1}{4}\text{Tr}\left\{  \rho_{0}+\sqrt{\rho_{1}}%
\sqrt{\rho_{0}}+\sqrt{\rho_{0}}\sqrt{\rho_{1}}+\rho_{1}\right\}  \right) \\
&  =-\log\left(  \frac{1}{2}\left(  1+\text{Tr}\left\{  \sqrt{\rho_{0}}%
\sqrt{\rho_{1}}\right\}  \right)  \right) \\
&  =\log\left(  \frac{2}{1+\text{Tr}\left\{  \sqrt{\rho_{0}}\sqrt{\rho_{1}%
}\right\}  }\right) \\
&  \geq\log\left(  \frac{2}{1+\sqrt{F\left(  W\right)  }}\right)  ,
\end{align*}
where the last line follows from%
\begin{align*}
\text{Tr}\left\{  \sqrt{\rho_{0}}\sqrt{\rho_{1}}\right\}   &  \leq
\text{Tr}\left\{  \left\vert \sqrt{\rho_{0}}\sqrt{\rho_{1}}\right\vert
\right\} \\
&  =\left\Vert \sqrt{\rho_{0}}\sqrt{\rho_{1}}\right\Vert _{1}\\
&  =\sqrt{F\left(  W\right)  }.
\end{align*}

The other inequality in (\ref{eq:rate-rel-2nd-bound}) follows from (21) in
Ref.~\cite{RFZ10}. In particular, they showed that%
\[
I\left(  W\right)  \leq H_{2}\left(  \frac{1}{2}\left(  1-\sqrt{F\left(
W\right)  }\right)  \right)  ,
\]
where the binary entropy $H_{2}\left(  x\right)  \equiv-x\log_{2}x-\left(
1-x\right)  \log_{2}\left(  1-x\right)  $. Combining this with the following
observation that holds for all $0\leq F\left(  W\right)  \leq1$ gives the
second inequality:%
\[
H_{2}\left(  \frac{1}{2}\left(  1-\sqrt{F\left(  W\right)  }\right)  \right)
\leq\sqrt{1-F\left(  W\right)  }.
\]

\end{IEEEproof}

\bigskip

\bigskip

\begin{IEEEproof}
[Proof of Proposition~\ref{prop:rate-cons-polarize}]These follow from the same
line of reasoning as in the proof of Arikan's Proposition~4 \cite{A09}. We
prove the first equality. Consider the mutual information%
\begin{align*}
I\left(  U_{1}U_{2};B_{1}B_{2}\right)   &  =I\left(  X_{1}X_{2};B_{1}%
B_{2}\right) \\
&  =I\left(  X_{1};B_{1}\right)  +I\left(  X_{2};B_{2}\right) \\
&  =2I\left(  W\right)  .
\end{align*}
By the chain rule for quantum mutual information~\cite{W11}, we have%
\begin{align*}
I\left(  U_{1}U_{2};B_{1}B_{2}\right)   &  =I\left(  U_{1};B_{1}B_{2}\right)
+I\left(  U_{2};B_{1}B_{2}U_{1}\right) \\
&  =I\left(  W^{-}\right)  +I\left(  W^{+}\right)  .
\end{align*}

The inequality follows because%
\begin{align*}
I\left(  W^{+}\right)   &  =I\left(  U_{2};B_{1}B_{2}U_{1}\right) \\
&  =I\left(  U_{2};B_{2}\right)  +I\left(  U_{2};B_{1}U_{1}|B_{2}\right) \\
&  =I\left(  W\right)  +I\left(  U_{2};B_{1}U_{1}|B_{2}\right)  .
\end{align*}
Thus,
\[
I\left(  W^{+}\right)  \geq I\left(  W\right)
\]
because $I\left(  U_{2};B_{1}U_{1}|B_{2}\right)  \geq0$~\cite{LR73,NC00,W11}%
.\ We then have%
\begin{align*}
2I\left(  W^{+}\right)   &  \geq2I\left(  W\right) \\
&  =I\left(  W^{-}\right)  +I\left(  W^{+}\right)  ,
\end{align*}
and the inequality follows.
\end{IEEEproof}

\bigskip

\bigskip

\begin{IEEEproof}
[Proof of Proposition~\ref{prop:rel-trans}]We begin with the first equality.
Consider that%
\begin{align*}
&  \sqrt{F\left(  W^{+}\right)  }\\
&  =\sqrt{F\left(  \rho_{0}^{+},\rho_{1}^{+}\right)  }\\
&  =\sqrt{F\left(  \frac{1}{2}\sum_{u_{1}}\left\vert u_{1}\right\rangle
\left\langle u_{1}\right\vert \otimes\rho_{u_{1}}^{B_{1}},\frac{1}{2}%
\sum_{u_{1}}\left\vert u_{1}\oplus1\right\rangle \left\langle u_{1}%
\oplus1\right\vert \otimes\rho_{u_{1}}^{B_{1}}\right)  }\\
&  \ \ \ \ \ \ \ \times\sqrt{F\left(  \rho_{0}^{B_{2}},\rho_{1}^{B_{2}%
}\right)  }\\
&  =\frac{1}{2}\left(  \sqrt{F\left(  \rho_{0}^{B_{1}},\rho_{1}^{B_{1}%
}\right)  }+\sqrt{F\left(  \rho_{1}^{B_{1}},\rho_{0}^{B_{1}}\right)  }\right)
\sqrt{F\left(  \rho_{0}^{B_{2}},\rho_{1}^{B_{2}}\right)  }\\
&  =F\left(  \rho_{0},\rho_{1}\right)  \\
&  =F\left(  W\right)
\end{align*}
The first two equalities follow by definition. The third equality follows from
the multiplicativity of fidelity under tensor product states~\cite{NC00,W11}:%
\[
F\left(  \rho\otimes\sigma,\tau\otimes\omega\right)  =F\left(  \rho
,\tau\right)  F\left(  \sigma,\omega\right)  .
\]
The fourth equality follows from the following formula that holds for the
fidelity of classical-quantum states:%
\begin{multline*}
\sqrt{F\left(  \sum_{x}p\left(  x\right)  \left\vert x\right\rangle
\left\langle x\right\vert \otimes\rho_{x},\sum_{x}p\left(  x\right)
\left\vert x\right\rangle \left\langle x\right\vert \otimes\sigma_{x}\right)
}\\
=\sum_{x}p\left(  x\right)  \sqrt{F\left(  \rho_{x},\sigma_{x}\right)  }.
\end{multline*}

We now consider the second inequality. The fidelity also has the following
characterization as the minimum Bhattacharya overlap between distributions
induced by a POVM\ on the states~\cite{FvG99,NC00,W11}:%
\[
F\left(  \rho_{0},\rho_{1}\right)  =\min_{\left\{  \Lambda_{m}\right\}
}\left(  \sum_{m}\sqrt{\text{Tr}\left\{  \Lambda_{m}\rho_{0}\right\}
\text{Tr}\left\{  \Lambda_{m}\rho_{1}\right\}  }\right)  ^{2}.
\]
So%
\[
\sqrt{F\left(  W^{-}\right)  }=\min_{\left\{  \Gamma_{m}^{B_{1}B_{2}}\right\}
}\sum_{m}\sqrt{\text{Tr}\left\{  \Gamma_{m}^{B_{1}B_{2}}\rho_{0}^{-}\right\}
\text{Tr}\left\{  \Gamma_{m}^{B_{1}B_{2}}\rho_{1}^{-}\right\}  }%
\]
Let $\Lambda_{m}$ denote the POVM that achieves the minimum for $\sqrt
{F\left(  W\right)  }$:%
\begin{align*}
\sqrt{F\left(  W\right)  }  & =\sqrt{F\left(  \rho_{0},\rho_{1}\right)  }\\
& =\min_{\left\{  \Lambda_{m}\right\}  }\sum_{m}\sqrt{\text{Tr}\left\{
\Lambda_{m}\rho_{0}\right\}  \text{Tr}\left\{  \Lambda_{m}\rho_{1}\right\}  }.
\end{align*}
Then the POVM\ $\left\{  \Lambda_{l}\otimes\Lambda_{m}\right\}  $ is a
particular POVM\ that can try to distinguish the states $\rho_{0}^{-}$ and
$\rho_{1}^{-}$. We then have%
\begin{align*}
&  \sqrt{F\left(  W^{-}\right)  }\\
&  \leq\sum_{l,m}\sqrt{\text{Tr}\left\{  \left(  \Lambda_{l}\otimes\Lambda
_{m}\right)  \left(  \rho_{0}^{-}\right)  \right\}  \text{Tr}\left\{  \left(
\Lambda_{l}\otimes\Lambda_{m}\right)  \left(  \rho_{1}^{-}\right)  \right\}
}\\
&  =\sum_{l,m}\sqrt{\text{Tr}\left\{  \left(  \Lambda_{l}\otimes\Lambda
_{m}\right)  \frac{1}{2}\left(  \rho_{0}^{B_{1}}\otimes\rho_{0}^{B_{2}}%
+\rho_{1}^{B_{1}}\otimes\rho_{1}^{B_{2}}\right)  \right\}  }\\
&  \ \ \ \ \ \times\sqrt{\text{Tr}\left\{  \left(  \Lambda_{l}\otimes
\Lambda_{m}\right)  \frac{1}{2}\left(  \rho_{1}^{B_{1}}\otimes\rho_{0}^{B_{2}%
}+\rho_{0}^{B_{1}}\otimes\rho_{1}^{B_{2}}\right)  \right\}  }%
\end{align*}%
\begin{multline*}
=\frac{1}{2}\sum_{l,m}\bigg[\bigg(\text{Tr}\left\{  \Lambda_{l}\rho_{0}^{B_{1}}\right\}
\text{Tr}\left\{  \Lambda_{m}\rho_{0}^{B_{2}}\right\}  +\\
\ \ \ \ \ \text{Tr}\left\{  \Lambda_{l}\rho_{1}^{B_{1}}\right\}
\text{Tr}\left\{  \Lambda_{m}\rho_{1}^{B_{2}}\right\}  \bigg)\bigg(\text{Tr}\left\{
\Lambda_{l}\rho_{1}^{B_{1}}\right\}  \text{Tr}\left\{  \Lambda_{m}\rho
_{0}^{B_{2}}\right\}  +\\
\ \ \ \ \ \ \text{Tr}\left\{  \Lambda_{l}\rho_{0}^{B_{1}}\right\}
\text{Tr}\left\{  \Lambda_{m}\rho_{1}^{B_{2}}\right\}  \bigg)\bigg]^{1/2}%
\end{multline*}
Making the assignments%
\begin{align*}
\alpha_{m} &  \equiv\text{Tr}\left\{  \Lambda_{m}\rho_{0}^{B_{2}}\right\}  ,\\
\beta_{l} &  \equiv\text{Tr}\left\{  \Lambda_{l}\rho_{0}^{B_{1}}\right\}  ,\\
\gamma_{l} &  \equiv\text{Tr}\left\{  \Lambda_{l}\rho_{1}^{B_{1}}\right\}  ,\\
\delta_{m} &  \equiv\text{Tr}\left\{  \Lambda_{m}\rho_{1}^{B_{2}}\right\}  ,
\end{align*}
the above expression is equal to%
\[
\sum_{l,m}\frac{1}{2}\sqrt{\beta_{l}\alpha_{m}+\gamma_{l}\delta_{m}}%
\sqrt{\gamma_{l}\alpha_{m}+\beta_{l}\delta_{m}}%
\]
We can then exploit Arikan's inequality in Appendix~D of Ref.~\cite{A09}\ to
have%
\begin{align*}
&  \sum_{l,m}\frac{1}{2}\sqrt{\beta_{l}\alpha_{m}+\gamma_{l}\delta_{m}}%
\sqrt{\gamma_{l}\alpha_{m}+\beta_{l}\delta_{m}}\\
&  \leq\sum_{l,m}\frac{1}{2}\left(  \sqrt{\beta_{l}\alpha_{m}}+\sqrt
{\gamma_{l}\delta_{m}}\right)  \left(  \sqrt{\gamma_{l}\alpha_{m}}+\sqrt
{\beta_{l}\delta_{m}}\right)  \\
&  \ \ \ \ \ \ \ -\sum_{l,m}\sqrt{\beta_{l}\alpha_{m}\gamma_{l}\delta_{m}}\\
&  =\sum_{l,m}\frac{1}{2}\left(  \alpha_{m}\sqrt{\beta_{l}\gamma_{l}}%
+\gamma_{l}\sqrt{\delta_{m}\alpha_{m}}+\beta_{l}\sqrt{\alpha_{m}\delta_{m}%
}+\delta_{m}\sqrt{\gamma_{l}\beta_{l}}\right)  \\
&  \ \ \ \ \ \ \ -\sum_{l}\sqrt{\beta_{l}\gamma_{l}}\sum_{m}\sqrt{\alpha
_{m}\delta_{m}}\\
&  =\sum_{l}\sqrt{\beta_{l}\gamma_{l}}+\sum_{m}\sqrt{\delta_{m}\alpha_{m}%
}-\sum_{l}\sqrt{\beta_{l}\gamma_{l}}\sum_{m}\sqrt{\alpha_{m}\delta_{m}}\\
&  =2\sqrt{F\left(  W\right)  }-F\left(  W\right)  .
\end{align*}

The inequality $F\left(  W^{-}\right)  \geq F\left(  W\right)  $ follows from
concavity of fidelity and its multiplicativity under tensor
products~\cite{NC00,W11}:%
\begin{align*}
  F\left(  W^{-}\right)  & =F\left(  \rho_{0}^{-},\rho_{1}^{-}\right)  \\
&  \geq\frac{1}{2}F\left(  \rho_{0}^{B_{1}}\otimes\rho_{0}^{B_{2}},\rho
_{1}^{B_{1}}\otimes\rho_{0}^{B_{2}}\right)  \\
&  \ \ \ \ \ +\frac{1}{2}F\left(  \rho_{1}^{B_{1}}\otimes\rho_{1}^{B_{2}}%
,\rho_{0}^{B_{1}}\otimes\rho_{1}^{B_{2}}\right)  \\
&  =\frac{1}{2}F\left(  \rho_{0}^{B_{1}},\rho_{1}^{B_{1}}\right)  F\left(
\rho_{0}^{B_{2}},\rho_{0}^{B_{2}}\right)  \\
&  \ \ \ \ \ +\frac{1}{2}F\left(  \rho_{1}^{B_{1}},\rho_{0}^{B_{1}}\right)
F\left(  \rho_{1}^{B_{2}},\rho_{1}^{B_{2}}\right)  \\
&  =\frac{1}{2}F\left(  \rho_{0}^{B_{1}},\rho_{1}^{B_{1}}\right)  +\frac{1}%
{2}F\left(  \rho_{1}^{B_{1}},\rho_{0}^{B_{1}}\right)  \\
&  =F\left(  W\right)
\end{align*}
The inequality $F\left(  W\right)  \geq F\left(  W^{+}\right)  $ follows from
the relation $\sqrt{F\left(  W^{+}\right)  }=F\left(  W\right)  $ and the fact
that $0\leq F\leq1$.
\end{IEEEproof}

\bibliographystyle{IEEEtran}
\bibliography{Ref}

\end{document}